\documentclass{mem}
\usepackage{natbib}\usepackage{txfonts}\usepackage{balance}
\usepackage{graphicx}
\usepackage[a4paper,breaklinks,dvipdfm]{hyperref}
\idline{00}{000}
\begin{document}

\title{SABAT: the ``Sapienza'' Balloon Trajectory\\Simulation Code}
\subtitle{}

\author{F.~G. Saturni\inst{1,2} 
\and R. Capuzzo-Dolcetta\inst{3}
\and P. de Bernardis\inst{3}
\and S. Masi\inst{3}}

\institute{
INAF -- Osservatorio Astronomico di Roma,
Via Frascati 33,
I-00040 Monte Porzio Catone (RM),
Italy
\and
ASI -- Space Science Data Center,
Via del Politecnico snc,
I-00133 Roma,
Italy
\and
``Sapienza'' University of Rome -- Dept. of Physics,
P.le A. Moro 5,
I-00185 Roma,
Italy\\
\email{francesco.saturni@inaf.it}
}

\authorrunning{Saturni}
\titlerunning{{\scriptsize SABAT} Simulation Code}

\abstract{We present {\scriptsize SABAT} (``Sapienza'' Balloon Trajectory Simulation Code), a Fortran software for the simulation of zero-pressure scientific balloon trajectories at stratospheric altitudes. Particular attention is made to the correct reproduction of the balloon ascending phase to the floating altitude. In order to obtain trajectories in agreement with those monitored during actual balloon launches and flights, {\scriptsize SABAT} features a balloon dynamical and geometric model, and a thermal model for the heat exchanges between the vehicle and the surrounding environment. In addition, both a synthetic description of the Earth's atmosphere and real data from radiosondes can be used. The validation tests give results in agreement with the characteristics of real balloon motion. Future developments of the code include optimization for balloons flying at polar latitudes, an improved treatment of the drag force acting on the balloon structure and the development of a user-friendly graphical interface.
}

\maketitle{}

\section{Introduction}

The problem of modeling the dynamical and thermal behavior of high-altitude, zero-pressure balloons dates back to early 1970s \citep[e.g.,][]{yaj09}. Efforts to systematically approach this problem in a multidisciplinary way led to the NASA's popular prediction software {\scriptsize THERMTRAJ} \citep{kre74,car81,hor83} and {\scriptsize SINBAD}\footnote{See also \citet{pan05} for {\scriptsize SINBAD}'s successor {\scriptsize NAVAJO}.} \citep{raq93}, which at present are however old and suffer from several limitations such as the inability to simulate 3D trajectories and the use of deprecated 1D atmospheric models. In order to overcome these limitation, new software has been developed, either by improving existing codes \citep[e.g.,][]{mus04,mus05} or by writing new ones \citep[e.g., the software {\scriptsize ACHAB} -- Analysis Code for High-Altitude Balloons developed by the CIRA -- Italian Aerospace Research Center in the framework of the PRORA-USV project;][]{pal07}.

In this paper we present {\scriptsize SABAT} (``Sapienza'' Balloon Trajectory Simulation Code), a Fortran code for the prediction of a zero-pressure stratospheric balloon trajectory. This software has been developed in the framework of the FATA project (PI: De Bernardis)\footnote{Project description available at {\ttfamily http://\\thermalvacuum.roma1.infn.it/simulazioni-\\numeriche-dello-sviluppo-di-traiettorie-\\di-sonde-a-quote-suborbitali/index.html}.}. The manuscript is organized as follows: in Sec. \ref{sec:sni}, we give an overview of the numerical treatment of a balloon floating at stratospheric altitudes in the Earth's atmosphere; in Sec. \ref{sec:cv}, we discuss the code validation; finally, in Sec. \ref{sec:conc} we summarize our results and illustrate the possible software updates.

\section{{\scriptsize SABAT} numerical implementation}\label{sec:sni}

{\scriptsize SABAT} treats a balloon during its flight as a thermodynamic system freely evolving inside a complex thermal environment subject to atmospheric winds. The full code works under the following assumptions:
\begin{itemize}
\item the balloon is considered a point mass with 3 degrees of freedom;
\item the gravity gradient over the balloon height and effects of humidity on atmospheric pressure are neglected;
\item the gaseous components are all assumed to follow the perfect gas law;
\item the lifting gas is considered transparent and of uniform pressure, temperature and density (except when valving or venting);
\item the balloon film temperature is assumed uniform along the surface.
\end{itemize}

\subsection{Balloon dynamical model}

\begin{figure}[htbp]
\begin{center}
\includegraphics[scale=0.25,angle=90]{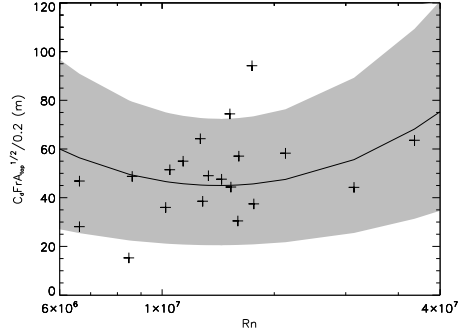}
\end{center}
\caption{Dimensionless drag coefficient as a function of the balloon Reynolds number. The best-fit relation according to Eq. \ref{eqn:drag} ({\itshape black solid line}), along with its 1$\sigma$ uncertainty ({\itshape grey shaded area}), is shown superimposed to the wind-tunnel balloon data by \citet[][{\itshape black crosses}]{fit74}.}
\label{fig:dcoeff}
\end{figure}

In the code, buoyancy is related to other forces acting on the balloon through a dynamical model. In order to compute a physically meaningful balloon trajectory, we assume that the balloon motion due to net buoyancy (also called gross inflation) $\mathbf{I}_{\rm gross} = \mathbf{g}\mathcal{V}\left( \rho_{\rm gas} - \rho_{\rm air} \right)$ for a balloon volume $\mathcal{V}$ and aerodynamic drag $\mathbf{D}$ is equivalent to a uniformly accelerated straight motion along all directions for short time intervals of integration. The acceleration $ {\ddot \mathbf{r}}$ is given by:
\begin{equation}\label{eqn:acc}
{\ddot \mathbf{r}} = \frac{\mathbf{g}\left[
\mathcal{V}\left(
\rho_{\rm gas} - \rho_{\rm air}
\right) + m_{\rm gross}
\right] + \mathbf{D}}{m_{\rm gross}+m_{\rm gas}+C_{\rm add}\left(
\rho_{\rm air}\mathcal{V}
\right)}
\end{equation}
Here, the gross mass $m_{\rm gross}$ is the overall mass of the non-gaseous components of the balloon, i.e. the sum of the payload mass, the ballast and the balloon film mass. The term $C_{\rm add}\left( \rho_{\rm air}\mathcal{V} \right)$ is the so-called added mass, which takes into account the airmass dragged by the balloon.

Following \citet{pal07}, we model the magnitude of the drag force $D$ as:
\begin{equation}\label{eqn:drag}
D = \frac{1}{2}\rho_{\rm air}v_{\rm rel}^2A_{\rm top}C_{\rm d},
\end{equation}
where $v_{\rm rel} = |\mathbf{v} - \mathbf{v}_{\rm wind}|$ is the balloon velocity relative to the wind speed, $A_{\rm top} \approx 1.086 \mathcal{V}^{1/3}$ is the balloon top projected area \citep{far05} and $C_{\rm d}$ is a drag coefficient that depends on the Reynolds number $Rn$, the Froude number $Fr$ and the balloon geometry \citep[see e.g.][and refs. therein]{dwy85}:
\begin{equation}\label{eqn:cd}
C_{\rm d} = 0.2 \frac{K_{\rm CD}}{Fr}\left(
\frac{k_1}{Rn} + k_2 Rn
\right)\frac{A_{\rm top}}{{A_{\rm top}^{(0)}}^{3/2}}
\end{equation}
We fit this relation to the wind-tunnel balloon data by \citet{fit74} to derive the free parameters $K_{\rm CD} = 19.8 \pm 1.2$ m, $k_1 = (1.50 \pm 0.78) \times 10^7$ and $k_2 = (0.86 \pm 0.43) \times 10^{-7}$ (see Fig. \ref{fig:dcoeff}).

\subsection{Geometric and thermal model}

The geometric model implemented in {\scriptsize SABAT} gives a description of the balloon shape, while the thermal model carefully considers the influence on the balloon of the surrounding environment in order to predict the flight performance. In fact, the balloon's vertical motion critically depends on the heat transfer between inner gas and balloon film \citep[see][for a detailed description]{far05}. The main assumptions at the basis of the thermal model are:
\begin{itemize}
\item the Sun is considered a blackbody at \hbox{5550 K};
\item Earth is considered as a gray body with given surface temperature and emittance;
\item Earth and atmosphere both emit in the infrared (IR);
\item the balloon film is considered as a gray body primarily emitting in the IR \hbox{($\sim$210 -- 270 K);}
\item heat sources are direct Sun radiation (day only), albedo (day only), diffuse IR radiation, internal IR self-glow, external and internal convection -- other possible heat sources are neglected.
\end{itemize}

\subsection{Atmospheric modeling}

A correct modeling of Earth's atmosphere is crucial in order to get a realistic description of the balloon trajectory. To this aim, the atmospheric model supplies information about air temperature, pressure and wind components. The script offers the possibility to choose between the International Standard Atmosphere \citep[ISA; e.g.,][]{tal75}, which samples Earth's atmosphere average conditions of pressure, temperature and density up to an altitude of 86,000 m (corresponding to a temperature of 186.87 K, or $-86.28$ $^\circ$C, and a pressure of 0.373 Pa), and daily measurements of atmospheric parameters obtained from radiosondes, that are launched from several stations in order to offer a worldwide coverage\footnote{Data publicly available e.g. at {\ttfamily http://\\weather.uwyo.edu/upperair/sounding.html}.}. In order to obtain a meaningful trajectory simulation, atmospheric data must be retrieved from the closest monitoring station to the balloon launching site.

\section{Code validation}\label{sec:cv}

With the features described above, {\scriptsize SABAT} is able to compute the 3D balloon trajectory starting from atmospheric conditions, balloon structure and time of launch as input. In order to validate the code, we have considered an extreme case of stratospheric balloon, with a volume of 340,000 m$^3$ and a total mass of \hbox{$\sim$4500 kg,} flying at polar latitudes from the site of Ny-{\AA}lesund (Svalbard Islands, Norway) at 78$^\circ$ 55$'$ 30$''$ N 11$^\circ$ 55$'$ 20$''$ E.

\subsection{Drag coefficient analysis}

We first analyzed the balloon ascending motion with different functional forms of the drag coefficient, depending on the balloon and wind speed moduli only. For this task, we adopted the ISA configuration as atmospheric model, and selected either:
\begin{enumerate}
\item Eq. \ref{eqn:cd} with an upper saturation value of 1.6 \citep{fit74};
\item Eq. \ref{eqn:cd} with no saturation;
\item a constant value $C_{\rm d} = 0.45$, which is typical of old balloon codes.
\end{enumerate}

\begin{figure}[htbp]
\begin{center}
\includegraphics[scale=0.3]{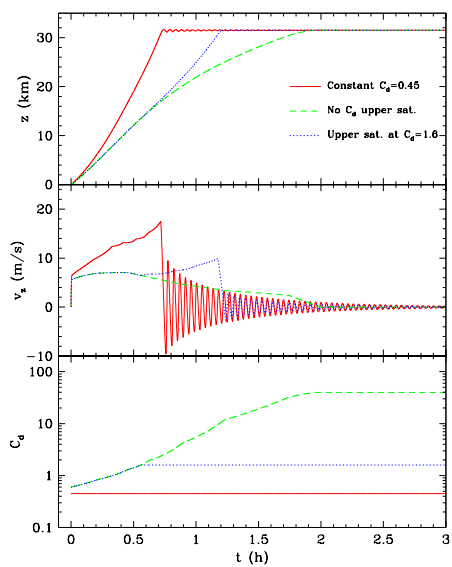}
\end{center}
\caption{Balloon vertical motion computed with different assumptions on the aerodynamic drag coefficient: constant drag ({\itshape red solid line}), saturating ({\itshape blue dotted line}) and non-saturating variable drag ({\itshape green dashed line}). {\itshape Upper panel:} balloon altitude. {\itshape Middle panel:} balloon vertical speed. {\itshape Lower panel:} drag coefficient as a function of time.}
\label{fig:ballalt}
\end{figure}

We show the comparison among the balloon vertical motions computed with each assumption in Fig. \ref{fig:ballalt}. A visual inspection reveals that a constant drag acting uniformly in all directions is unsuitable to correctly treat the balloon ascension, since the system reaches the floating altitude in a rather short time of $\sim$0.8 h with a maximum speed of $\sim$17 m s$^{-1}$. Conversely, the vertical motion evaluated with a non-saturating aerodynamic drag produces an overestimated atmospheric resistance that excessively damps the balloon ascension and oscillations around the floating altitude. Therefore, we finally selected the saturating drag coefficient as the most suitable description for the action of drag force on the balloon.

\begin{figure}[htbp]
\begin{center}
\includegraphics[scale=0.3]{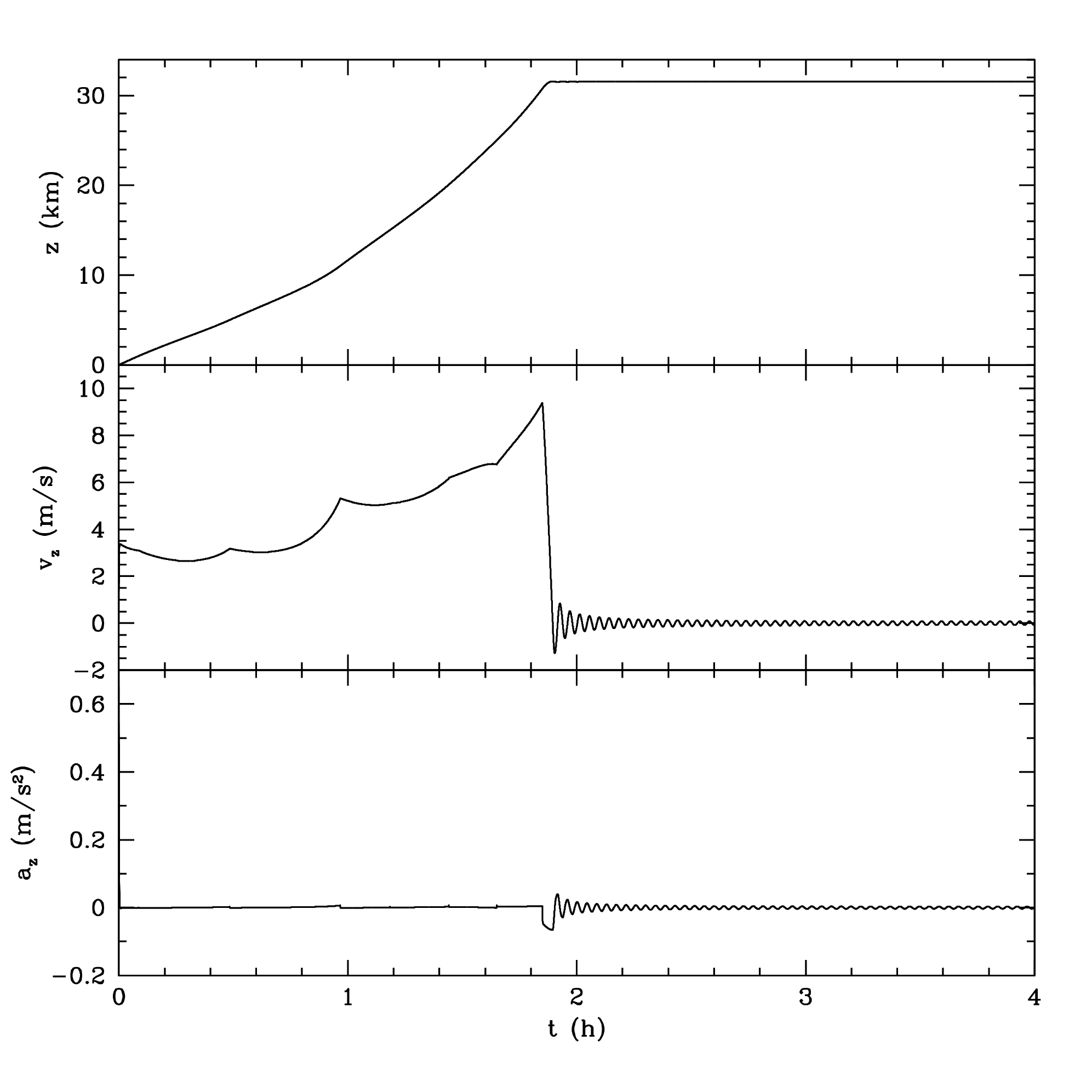}
\end{center}
\caption{{\itshape Top panel:} balloon ascending motion to the floating altitude. {\itshape Middle panel:} balloon vertical speed dependence on time. {\itshape Bottom panel:} balloon vertical acceleration due to net buoyancy.}
\label{fig:dragspec}
\end{figure}

Subsequently, we specialized the drag coefficient functional form acting along a specific direction that depends only on the balloon and wind speed components in that direction (see Fig. \ref{fig:dragspec}). In this way, we obtained a balloon ascending in $\sim$1.9 h to a floating altitude of $\sim$32 km, with a maximum ascending speed of $\sim$10 m s$^{-1}$.

\subsection{Simulations of balloon ascension with actual atmospheric data}

\begin{figure}[htbp]
\begin{center}
\includegraphics[scale=0.3]{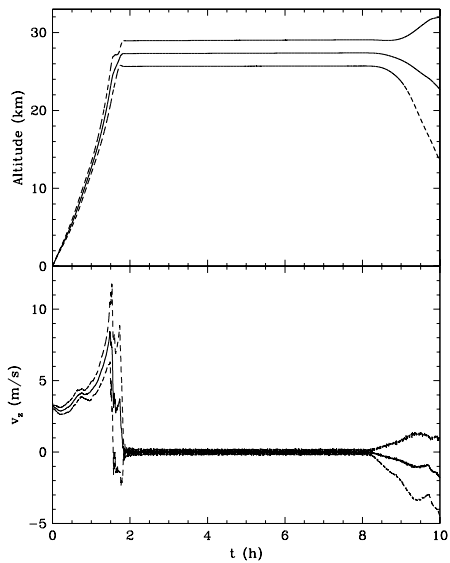}
\end{center}
\caption{Balloon flight simulation with real atmospheric data from radiosonde sampling. {\itshape Upper panel:} average balloon altitude ({\itshape solid line}) with 1$\sigma$ confidence intervals ({\itshape dashed lines}). {\itshape Lower panel:} average balloon vertical speed (solid line) with 1$\sigma$ confidence intervals ({\itshape dashed lines}).}
\label{fig:zsim}
\end{figure}

The last step of the code development was focused on simulating ``real'' balloon flights from actual atmospheric input data. In order to accomplish this task, we used the sounding data for a specified date and time of operations. We therefore ran a 10 h balloon-flight simulation starting on December 15 at 12:00 UT for the years from 2003 to 2016, discarding runs in which the balloon lands before 10 h of flight.

Fig. \ref{fig:zsim} shows the resulting simulation for a total of 11 winter flights. The balloon rises up to a floating altitude of $27.3 \pm 1.7$ km in $\sim$2 h with a maximum vertical speed of $8.4^{+3.3}_{-2.3}$ km s$^{-1}$, and lasts floating for $\sim$6 h during which the gas losses due to venting gradually reduce the balloon buoyancy. Peculiar atmospheric conditions may extend the balloon floating phase in time,which produces the large spread of up to $\sim$18 km in the final altitude during the balloon fall-down.

\section{Conclusions}\label{sec:conc}

In this paper, we have presented the Fortran code {\scriptsize SABAT} for simulations of atmospheric balloon trajectories. Our script is able to reproduce the vertical motion of a zero-pressure balloon. The preliminary study on the balloon ascending phase pointed out that:
\begin{itemize}
\item the atmospheric drag force is best modeled with a variable drag coefficient depending on the balloon and wind speed in each direction, saturating at a fixed value;
\item simulating the balloon flight for different real atmospheric conditions, the balloon reaches a floating altitude of $\sim$27 km ascending in $\sim$2 h with a maximal vertical velocity of $\sim$8 m s$^{-1}$, lasting for $\sim$6 h before starting to fall due to venting gas losses. The 1$\sigma$ spread in floating altitudes is $\sim$4 km, which increases up to $\sim$18 km in the fall-down phase depending on the atmospheric conditions.
\end{itemize}

A future extension of the balloon trajectory simulations to vehicles specifically designed for long-duration flights at polar latitudes is planned. Future versions of this code for balloon trajectory simulations will also feature: (i) a better treatment of the aerodynamic drag in order to verify possible overestimates of the drag force \citep[e.g.,][]{pal07}, and (ii) the development of a user interface in order to allow the code to be used with simplified input, running and output operations.

\begin{acknowledgements}
We thank I. Musso (Altec SPA) and R. Palumbo (CIRA) for useful discussion. We acknowledge funding from Progetto ``FATA'' Regione Lazio FILAS RU-2014-1038.
\end{acknowledgements}

\bibliographystyle{mem}
\bibliography{sbtsc}

\end{document}